# Extreme-ultraviolet spatiotemporal vortices via high harmonic generation

Rodrigo Martín-Hernández[1,2*], Guan Gui[3], Luis Plaja[1,2], Henry K. Kapteyn[3], Margaret M. Murnane[3], Chen-Ting Liao[3, 4*], Miguel A. Porras[5], and Carlos Hernández-García[1,2]

[1]Grupo de Investigación en Aplicaciones del Láser y Fotónica, Departamento de Física Aplicada, Universidad de Salamanca, 37008, Salamanca, Spain
[2]Unidad de Excelencia en Luz y Materia Estructuradas (LUMES), Universidad de Salamanca, Salamanca, Spain
[3]JILA and Department of Physics, University of Colorado and NIST, Boulder, CO 80309, United States of America
[4]Department of Physics, Indiana University, Bloomington, IN 47405, United States of America
[5]Complex Systems Group, ETSIME, Universidad Politécnica de Madrid, Ríos Rosas 21, 28003 Madrid, Spain

*liao3@iu.edu
*rodrigomh@usal.es

**Abstract**

Spatiotemporal optical vortices (STOV) are space-time structured light pulses with a unique topology that couples spatial and temporal domains and carry transverse orbital angular momentum (OAM). Up to now, their generation has been limited to the visible and infrared regions of the spectrum. During the last decade, it was shown that through the process of high-order harmonic generation (HHG) it is possible to up-convert spatial optical vortices that carry longitudinal OAM from the near-infrared into the extreme-ultraviolet (EUV), thereby producing vortices with distinct femtosecond and attosecond structure. In this work we demonstrate theoretically and experimentally the generation of EUV spatiotemporal and spatiospectral vortices using near infrared STOV driving laser pulses. We use analytical expressions for focused STOVs to perform macroscopic calculations of HHG that are directly compared to the experimental results. As STOV beams are not eigenmodes of propagation, we characterize the highly-charged EUV STOVs both in the near and far fields, to show that they represent conjugated spatiotemporal and spatiospectral vortex pairs. Our work provides high-frequency light beams topologically coupled at the nanometer/attosecond scales domains with transverse OAM, that could be suitable to explore electronic dynamics in magnetic materials, chiral media, and nanostructures.

## 1   Introduction

Structured light fields, tailored in their spatial, spectral, polarization and temporal properties, represent fundamental light science with a rich and important range of applications in imaging and spectroscopy [1,2]. In the past decade, a family of spatiotemporal coupled structured laser beams called spatiotemporal optical vortices (STOV) has attracted attention due to their unique space-time inseparable topological properties. The topology of light is typically associated with the presence of a spatial phase singularity in a vortex beam. In the 1990's, the spiral phase profile carried by vortex beams was associated with a longitudinal orbital angular momentum (OAM) [3], which was found quantized by the topological charge $l$ —or the number of $2\pi$-phase jumps along the azimuthal dimension— even at a single photon level. Since then, longitudinal OAM beams have been increasingly used in many applications including quantum optics, optical communications, imaging, metrology, and ultrafast science [4–8].

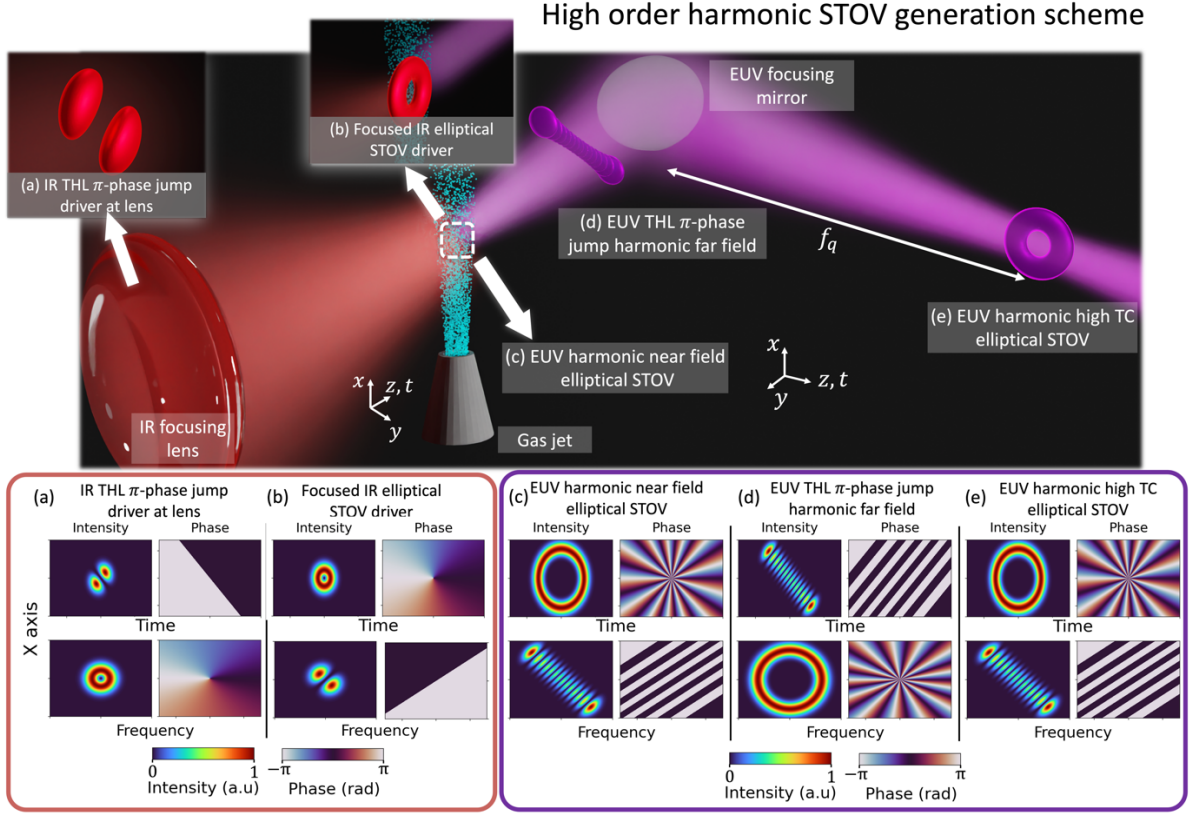

Figure 1: High-order spatiotemporal harmonic vortex generation setup. (a) An IR Hermite-lobed (THL) driver beam is focused, generating an elliptical STOV distribution, (b), in a gas target. (c) HHG upconverts the IR STOV into EUV high topological charge, elliptical harmonic STOVs. (d) Due to the STOV's propagation dynamics, the high topological charge elliptical near-field STOV evolves into a spatiotemporal lobulated intensity profile in the far-field. However, in the spatiospectral domain, an elliptical intensity distribution is observed. (e) An EUV focusing mirror is used to refocus the far-field emission, resulting in an elliptical EUV high-topological charge STOV at the mirror's focus. Note that results in panels (a) to (e) are based on analytical and numerical solutions used for illustrative purposes.

While longitudinal vortex beams are characterized by their spiral phase profile in the transverse spatial plane, STOVs exhibit a twisted phase profile in the spatiotemporal plane. Thus, the topological charge of STOVs refer to the number of $2\pi$-phase jumps in the plane that combines the temporal and one of the transverse spatial dimensions. As such, STOVs are also known as spatiotemporal OAM (ST-OAM) beams and transverse OAM beams, i.e., vortex beams that carry OAM along the direction transverse to propagation. STOVs were theoretically described about two decades ago [9,10]. Since their first experimental observation in a nonlinear filamentation process [11], several groups [12–14] generated, almost simultaneously, STOVs in the near infrared (IR) or visible regions using standard optics, such as 4f pulse shapers incorporating spiral or $\pi$-step phase plates [12], or spatial light modulators [13,15]. More recently, theories and simulations suggest the generation of STOV by photonic crystal slabs [16] or metasurfaces [17], and experiments realized them with meta-optics [18]. The experimental generation, characterization [12,13,19,20] and recent applications as, e.g., information carriers [15], have encouraged theoretical work [21–26] to understand their propagation properties and transverse OAM. Using perturbative second harmonic generation, near-IR STOVs have been up-converted into the visible spectral regime, with a corresponding doubling of their topological charge [27,28].

By exploiting the extreme non-perturbative non-linear process of high-order harmonic generation (HHG), the generation of high-frequency EUV, longitudinal OAM vortices have been demonstrated. In HHG [29,30], an intense femtosecond IR vortex beam is focused into a gas, producing a comb of harmonics with comparable intensities extending up to a maximum photon energy that scales linearly with the driver intensity and quadratically with its central wavelength [31,32]. From the macroscopic point of view, the coherent addition of the harmonic radiation emitted from all atoms within the gas medium needs to be considered to enable phase-matching, so that bright spatially-coherent beams can be generated [33,34]. In the case of HHG driven by single-mode IR longitudinal vortices, thanks to OAM conservation, the topological charge scales linearly with the harmonic order, and highly charged EUV vortices are generated [35–37]. In fact, the versatility and high coherence of the HHG process makes it possible to produce a variety of EUV beams that are spatially structured in their phase, amplitude and/or polarization properties by carefully engineering the driving laser field. This includes the generation of EUV vector beams [38], circularly polarized highly charged harmonic vortex beams [39], low-charge harmonic vortex beams [40,41], EUV beams with time-dependent OAM or self-torque [42,43], low divergence EUV harmonic combs [44], or attosecond vortex pulse trains [45], among others. Moreover, very recently OAM EUV high-harmonic beams were used to demonstrate non-destructive low-dose imaging of highly-periodic and delicate structures, with far less damage than in the case of electron microscopy [46].

Spatiotemporal coupling in HHG has enabled the creation of unique high harmonic and attosecond pulses by using spatially chirped IR driving pulses. For instance, the use of transversely chirped focused driving beams—where the local frequency varies along the focal spot—lead to the attosecond lighthouse effect [47], allowing subsequent attosecond pulses to propagate in different spatial directions. This effect has been harnessed for generating spatially isolated attosecond pulses [48–50]. A related outcome, where an ultrafast wavefront emerged in the harmonic beam, can be obtained through a noncollinear driving scheme [51]. In addition, using driving beams with angular spatial chirp at the focus—a scheme known as simultaneous space-time focusing [52] —enable the generation of spatially chirped high-order harmonics [53].

Consequently, the high degree of spatiotemporal control provided by HHG makes it an ideal candidate for generating high-frequency STOVs. Indeed, it has been recently theoretically proposed that HHG could be used to generate highly-charged EUV STOVs [54], though critical aspects, such as phase-matching, far-field propagation dynamics and EUV refocusing, and its experimental realization remain unexplored. The complex behaviour of STOVs, which, in contrast to longitudinal vortex beams, are not eigenmodes of propagation, introduces additional challenges, increasing the complexity of their generation and potential applications.

In this work, we demonstrate theoretically and experimentally the generation of EUV STOVs that carry highly-charged transverse OAM through the process of HHG. We note the duality in the focusing dynamics of STOVs, whereby they are transformed into spatiospectral optical vortices (SSOVs) and vice versa. As such, high harmonic STOVs and SSOVs are short-wavelength space-time vortices that feature elliptical intensity and spiralling phase, both in their respective spatial-temporal and spatial-spectral planes. Through focusing an IR tilted Hermite-lobed driving field into a gas (see Fig. 1), we up-convert its spatiotemporal properties to generate highly charged EUV SSOVs, which, upon proper focusing can be converted into highly charged EUV STOVs. We demonstrate that the topological charge of the harmonic vortices (either in their spatiotemporal or spatiospectral versions) scales linearly with that of the driving laser field, and we show how the properties of EUV STOV and SSOVs can be controlled by tailoring the driving laser field. The excellent agreement between experiment and theory allows us to characterize the properties of highly charged harmonic STOVs and SSOVs.

## 2    Results

### 2.1    Spatiotemporal and spatiospectral optical vortex propagation dynamics

Since STOVs are not eigenmodes of free-space propagating wave equations, a careful analysis of their propagation dynamics is crucial for understanding the results of the HHG STOV experiments. Fields that focus to STOVs have been studied by numerical means and experimentally in [21,22] and extended to high numerical apertures and short durations in [55]. As demonstrated by pure analytical means in [56], a STOV of arbitrary topological charge is produced when focusing the spatiotemporal, Gaussian-enveloped, tilted Hermite-lobed (THL field):

$$E(x,y,t,0) = e^{-\frac{y^2}{\sigma_y^2}} e^{-\frac{x^2}{\sigma_x^2}} e^{-\frac{t^2}{\sigma_t^2}} H_l\left(\frac{t}{\eta\sigma_t} \pm \frac{x}{\sigma_x}\right) e^{-i\omega_0 t}, \tag{1}$$

where $\omega_0$ is the central angular frequency, $\sigma_x$, $\sigma_y$ and $\sigma_t$ are the $1/e^2$ widths in $x$, $y$ and $t$, $\eta$ is an "inhomogeneity" parameter, and the order $l$ of the Hermite polynomial $H_l$ determines the number of $\pi$-steps in the phase between the $l+1$ tilted lobes. Fig. 1 (a) shows the intensity and phase profiles in time and frequency of a THL with $l=1$. In our HHG configuration, an IR THL field is created with a non-cylindrically symmetric 4-f pulse shaper (see Methods), and then focused to yield an IR STOV at the gas jet. The focused field at any propagation distance $z$ reads [56]:

$$E(x,y,t',z) = \left[\frac{q_{y,0}}{q_y(z)}\right]^{\frac{1}{2}} e^{\frac{ik_0 y^2}{2q_y(z)}} \left[\frac{q_{x,0}}{q_x(z)}\right]^{\frac{1}{2}} e^{\frac{ik_0 x^2}{2q_x(z)}} e^{-\frac{t'^2}{\sigma_t^2}} \left[\left(1 - \frac{z}{f}\right) \frac{q_{x,0}}{q_x(z)}\right]^{\frac{l}{2}}$$

$$\times H_l\left\{\left[\frac{q_x(z)}{q_{x,0}} \frac{1}{\left(1-\frac{z}{f}\right)}\right]^{\frac{1}{2}} \left(\frac{t'}{\eta\sigma_t} \pm \frac{x}{\sigma_x} \frac{q_{x,0}}{q_x(z)}\right)\right\} e^{-i\omega_0 t'}, \tag{2}$$

where $k_0 = \frac{\omega_0}{c}$, $t' = t - \frac{z}{c}$ is the local time, $\frac{1}{q_{\alpha,0}} = -\frac{1}{f} + \frac{2i}{k_0 \sigma_\alpha^2}$ ($\alpha = x,y$), $q_\alpha(z) = z + q_{\alpha,0}$, and $f$ is the focal length. At the focal plane, $z = f$, the field

$$E(x,y,t',f) \propto e^{-\frac{y^2}{\sigma_{y,f}^2}} e^{-\frac{x^2}{\sigma_{x,f}^2}} e^{-\frac{t'^2}{\sigma_t^2}} \left(\frac{t'}{\eta\sigma_t} \mp i\frac{x}{\sigma_{x,f}}\right)^l e^{-i\omega_0 t'}, \tag{3}$$

with $\eta = 1$ is a canonical, elliptical STOV of focal width $\sigma_{\alpha,f} = \frac{2f}{k_0}\sigma_\alpha$, where $\alpha = x, y$. Fig. 1(b) shows the intensity and phase profiles in time and frequency of the STOV at $z = f$ with $l = 1$. An inhomogeneity parameter $\eta$ different from unity accounts for imperfect, non-elliptical STOVs whose duration or "width" $\sigma_t$ does not match the vortex core duration $\eta\sigma_t$ and will be considered later. Defining a spatiotemporal azimuthal angular coordinate $\zeta = \arctan\left(\frac{x\sigma_t}{t'\sigma_{x,f}}\right)$, the azimuthal phase profile $\phi$ of the canonical STOV is $\phi(x,t') = \mp l\zeta$, which amounts a phase profile $\pm l\zeta$ in $(x,z)$ corresponding to a topological charge $\pm l$ of the transverse phase line singularity.

At the far field ($z \to \infty$), the field described by Eq. (2) becomes again a spatiotemporal THL field,

$$E(\theta_x, \theta_y, t') \propto e^{-\frac{\theta_x^2}{\theta_{x,0}^2}} e^{-\frac{\theta_y^2}{\theta_{y,0}^2}} e^{-\frac{t'^2}{\sigma_t^2}} H_l\left(\frac{t'}{\eta\sigma_t} \mp \frac{\theta_x}{\theta_{x,0}}\right) e^{-i\omega_0 t'}, \tag{4}$$

similar in shape to that of Fig. 1(a), but in the far-field observation angle variables $\theta_\alpha = \frac{\alpha}{z}$ ($\alpha = x,y$) and divergence angles $\theta_{\alpha,0} = \frac{\sigma_\alpha}{f}$.

Interestingly, the spatiospectrum $\hat{E}$, or temporal Fourier transform of $E$, follows an inverted dynamic. The spatiospectrum of the spatiotemporal THL field on the lens has the shape of a STOV but in spatiospectral domain [Fig. 1(a) for $l=1$], which we denote as SSOV of topological charge $\mp l$. The spatiospectrum of the canonical STOV at $z = f$ is THL shaped with $l$ $\pi$-steps in the phase [Fig. 1(b)], and the spatiospectrum of the spatiotemporal THL far field is again a SSOV of opposite topological charge $\pm l$ (not shown). Reversal of the topological charge is a known phenomenon in STOVs [25] and affects also to SSOVs. Thus, there is an alternating duality between STOVs and SSOVs in their respective domains upon focusing and propagation up to the far field. In fact, it was recently found that when a STOV loses its symmetry, the intrinsic OAM is preserved but the STOV may ambiguously feature spatiotemporal phase singularities with topological charge of equal or opposite signs [25].

## 2.2 Generation of high-topological charge spatiotemporal and spatiospectral optical vortices in the EUV

By focusing a THL field [see Fig.1 (a)], an elliptical IR STOV is created at the focus [see Fig.1 (b)], driving HHG in a low-density gas jet. As a result of the highly nonlinear up-conversion process, high harmonic STOVs emerge at the target plane [Fig. 1 (c)]. Upon propagation to the far field, EUV STOVs evolve into a lobulated spatiotemporal intensity distribution [Fig. 1 (d)], which presents a spatially resolved spectrum shaped as SSOVs. These spatiospectral harmonic vortices can be successively transformed into harmonic STOVs by using an EUV focusing mirror [Fig. 1 (e)].

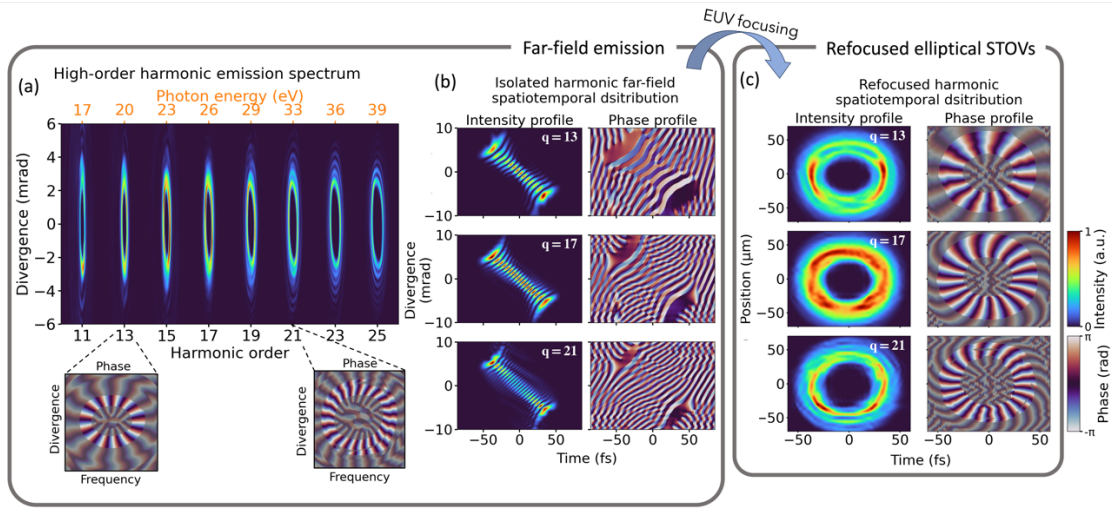

Figure 2: Simulation results for the generation of EUV high harmonic STOVs and SSOVs. (a) STOV driven HHG spectrum. The insets show the spatiospectral helical phase distribution for the 13th and 21st harmonic orders. (b) Spatiotemporal intensity (left column) and phase (right column) far-field distribution for harmonic orders 13th, 17th and 21st. (c) Spatiotemporal intensity (left column) and phase (right column) of the refocused far-field emission for the harmonic orders in panel (b), assuming a 20 cm focal length focusing mirror, leading to a well-defined transverse OAM beam.

A first approach to identify the up-conversion of STOVs in HHG consists in analyzing the nonlinear mechanism from a simple model based on the strong field approximation [57,58] (see Methods). The $q$-th harmonic order near field can be described by $E_q^{nf} = |E_0|^p e^{i(q\phi_0 + \phi_{int})}$, where $E_0$ and $\phi_0 = \mp l_0 \zeta$ are the amplitude and spatiotemporal phase of an elliptical STOV driving field of topological charge $l_0$ described by Eq. (3), the $p$-th power corresponds to the HHG process non-perturbative intensity scaling, and $\phi_{int}$ is the intrinsic dipole phase [59,60]. As a first approximation where the intrinsic phase is neglected—an approach that has been explored in HHG driven by longitudinal vortex or vector beams [58,61]—, the harmonic spatiotemporal phase scales linearly with the harmonic order as $\phi_q \simeq q\phi_0 = \mp q l_0 \zeta$, as depicted in Fig. 1(c). Thus, the topological charge of the $q$-th harmonic STOV scales as $l_q = q l_0$, as predicted in [54], and mimics the up-conversion law for the topological charge of longitudinal vortex beams [35]. In addition, the amplitude, $|E_0|^p$, of the $q$-th harmonic near field retains the elliptical symmetry of the driver, as depicted in Fig. 1 (c). Therefore, this simple model of HHG driven by STOV fields predicts the emergence of $l_q$-order elliptical STOVs at the near field, whose signature at the far field is a spatiotemporal THL field featuring $l_q$ $\pi$-steps between $l_q + 1$ lobes, or equivalently, $l_q$-order SSOVs of topological charge $l_q$, as illustrated in Fig. 1 (d). Given the SSOV-STOV duality, the far-field radiation can be refocused with an EUV focusing mirror to obtain highly charged elliptical STOVs at the focus of the mirror, as depicted in Fig. 1(e).

In order to fully analyze the properties of the high-order harmonics driven by an IR STOV, we have performed advanced quantum HHG simulations, that include both the single-atom and macroscopic response using the electromagnetic field propagation [62], a method that has proven to be extremely successful when describing and predicting HHG driven by structured light beams (see Methods). Note that these simulations include the dipole intrinsic phase that was previously neglected in the simple model. In Figs. 2(a) and 2(b) we show the resulting far-field harmonic spatiospectral and spatiotemporal (strictly speaking, angular-temporal) distributions obtained in an atomic Hydrogen target driven by a canonical STOV with the same transverse OAM, $l_0 = 1$, at the focus. The IR STOV at 800 nm central wavelength is modelled by a $\sim 52$ fs FWHM long sin$^2$ temporal envelope, $1.6\times10^{14}$ W/cm$^2$ peak intensity.

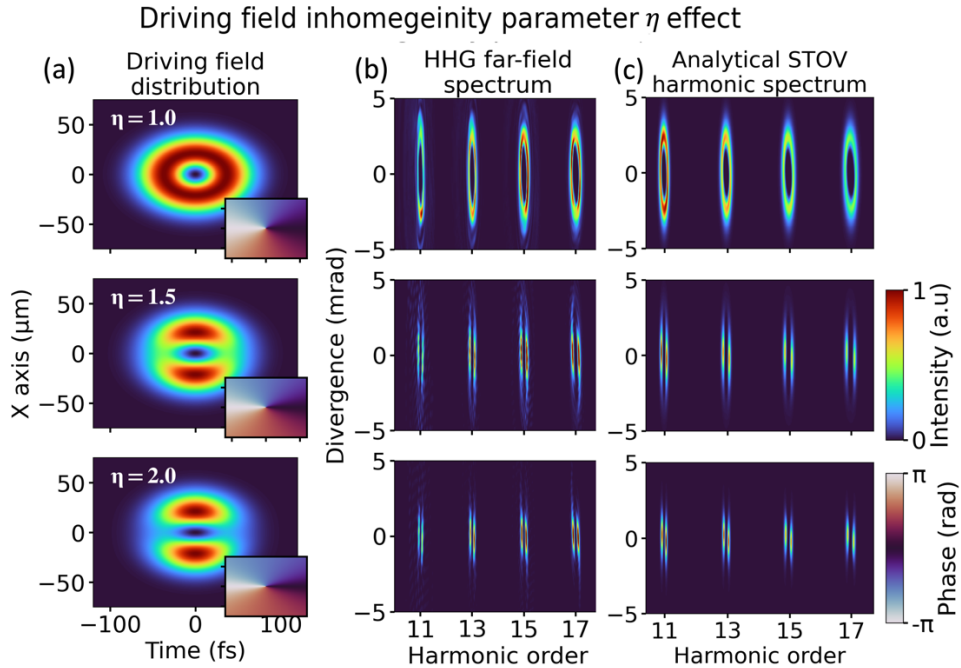

Figure 3: Influence of the inhomogeneity parameter in the generation of EUV STOVs given the same transverse OAM $l_0 = 1$. (a) Simulated driving IR STOV spatiotemporal intensity distribution for different inhomogeneity parameters $\eta$. The inset shows the spatiotemporal phase profile in each case. (b) Full quantum simulation with far field propagation HHG spatiospectral intensity profiles using the IR driving STOVs shown in panel (a). (c) Analytical far-field spatiospectra obtained propagating the harmonic spatiotemporal near field in the strong field approximation up to the far field and performing a temporal Fourier transform (see Methods).

Figure 2 (a) presents the spatiospectral distribution of the simulated far-field harmonics. Each odd-order harmonic exhibits the elliptical intensity distribution of a SSOV, corresponding to a STOV in the near field. The two insets show the spatiospectral phase distributions for the 13th and 21st harmonic orders, which present twisted phase profiles with the corresponding topological charges $l_{13} = 13$ and $l_{21} = 21$, confirming that the harmonics are emitted at the far-field as highly charged SSOVs. Interestingly, higher harmonic orders show broader spectra as a result of the increasing radius $\propto \sqrt{l_q}$ of SSOVs (similarly to STOVs) with the linearly increasing topological charge.

The spatiotemporal distribution of far-field harmonics is obtained through spectral filtering and subsequent inverse temporal Fourier transform. In Fig. 2(b), we present the spatiotemporal distributions of the 13th, 17th and 21st harmonic orders. As expected, the far-field distributions present the THL structure with $l_q + 1$ lobes and $l_q$ phase $\pi$-steps associated with the $l_q$-charged STOV at the focus—or the $l_q$-charged SSOV at the far-field.

In order to obtain high-topological charge EUV canonical STOVs that can be used in experiments, further refocusing transforms SSOV to STOV, as illustrated in Fig. 1 (e). Specifically, we perform Fresnel propagation of the isolated harmonic orders in Fig. 2(b) after 10 cm of propagation from the gas jet using a $f_q = 20$ cm focal length (see Supplementary Information). The results presented in Fig. 2 (c), demonstrate the generation of high-topological charge, elliptical EUV harmonic STOVs after refocusing the harmonics driven by a canonical IR elliptical STOV at focus. The phase profile for the 13th, 17th and 21st harmonic orders clearly show the linear scaling of the topological charge, now in the spatiotemporal domain.

From an experimental perspective, generating a perfectly circularly symmetric canonical IR driving STOV in space-time at the focus is challenging. Imperfect STOVs can be modelled by an inhomogeneity parameter $\eta$ differing from unity—see Eq. (3). Figure 3 compares the far-field high harmonic SSOVs when they are driven by the canonical STOVs (first row) or imperfect STOVs modelled with $\eta = 1.5$ (second row) and $\eta = 2$ (third row) at the focal plane. Increasing $\eta$ results in an intensity profile redistribution of the driver in the spatiotemporal plane that deforms the elliptical intensity towards two strips, while the phase profiles in the insets remain almost unaltered. This indicates that asymmetric STOVs, geometrically speaking, do not really affect transverse OAM when considering topology. Figures 3(b) and 3(c) show the high harmonic spectra from the numerical simulations and from the analytical model based on the strong field approximation, respectively. In either case, inhomogeneity of the driver translates into imperfect high-order harmonic SSOVs, whose elliptical intensity is structured into two strips. The excellent agreement between the full numerical simulations and the analytical model underscores the minor role played here by the intrinsic harmonic phase.

## 2.3 Experimental EUV STOVs via HHG.

Once the properties of high-order harmonic STOVs have been analyzed, we present the experimental results that confirm our predictions. In the experiment, a STOV beam with $l_0 = 1$ is generated by a 4-f pulse shaper serving as the driving laser for HHG. A Xenon gas jet is placed around the focal plane of the driving laser beam, where an EUV beam is generated through HHG process. The EUV beam is characterized by an imaging spectrometer at the far-field. The experimental HHG-STOV setup is sketched in Fig. 4 (a), where an IR Gaussian beam is sent to a STOV pulse shaper and subsequently directed to the HHG line where it is focused into a STOV at the Xenon gas cell. We note that the driving beam for HHG is an elliptical STOV with inhomogeneity parameter $\eta = 1.8$ (close to $\eta = 2$) due to an elliptical beam profile in the pulse shaper. Therefore, we compare our experimental results with simulations for the case of $\eta = 2$. More detailed discussions on the inhomogeneity parameter can be found in the Supplementary Information.

As evidenced in the previous section, high-harmonic STOVs are highly sensitive to propagation. In Fig. 4 we show the experimental and theoretical results of HHG driven by elliptical STOVs for different relative positions between the gas jet and the focus —before the STOV focus ($z = f - 1.5\ mm$), at the STOV focus ($z = f$) and after the STOV ($z = f + 1.5\ mm$), where $z$ is the gas jet location with respect to the STOV focus. As in Fig. 2(a), the spectral bandwidth increases with the harmonic order, indicative of the topological charge scaling with the harmonic order. The observed double striped harmonic intensity distributions agree with the inhomogeneous elliptical STOV driver case previously discussed in Figs. 3(a)-(c). However, in the case of the out-of-focus measurements ($z = f - 1.5\ mm$ and $z = f + 1.5\ mm$), the intensity fringes are not perfectly on-axis aligned, and the line connecting the stripes' maxima points (dashed lines) clearly show a z-dependent slope. Starting with a positive slope when the gas jet is located before the STOV focus, it evolves into a negative slope if the gas jet is placed after the STOV focus, while it tends to zero close to the STOV focus. We note that these observations are in excellent agreement with the numerical results and the analytical model. The sign flip behaviour can be explained by the driver's intensity profile tilt-change suffered by the focused IR STOV as it propagates and passes through the focus, as shown in Fig. 4 (a) insets (note the white dashed lines in each case). Due to the large inhomogeneity ($\eta > 1$) in the driving STOV, at focus (middle row in Fig. 4), we can clearly distinguish the two lobes opposed along the spatial (vertical) axis. On the contrary, before- and after-the-focus positions, — corresponding to first and third row in Fig. 4—, the two lobes in the driver STOV are not aligned. Again, the slope of the line connecting the lobes' maximum intensity points swaps the sign as it goes from before- to after-the-focus positions. As the HHG process translates the properties of the driving beam to the high-order harmonics, together with the propagation properties of STOVs, the slope sign flip in the temporal domain is captured as a slope sign swap for the stripes observed in the spectral domain for the HHG emission.

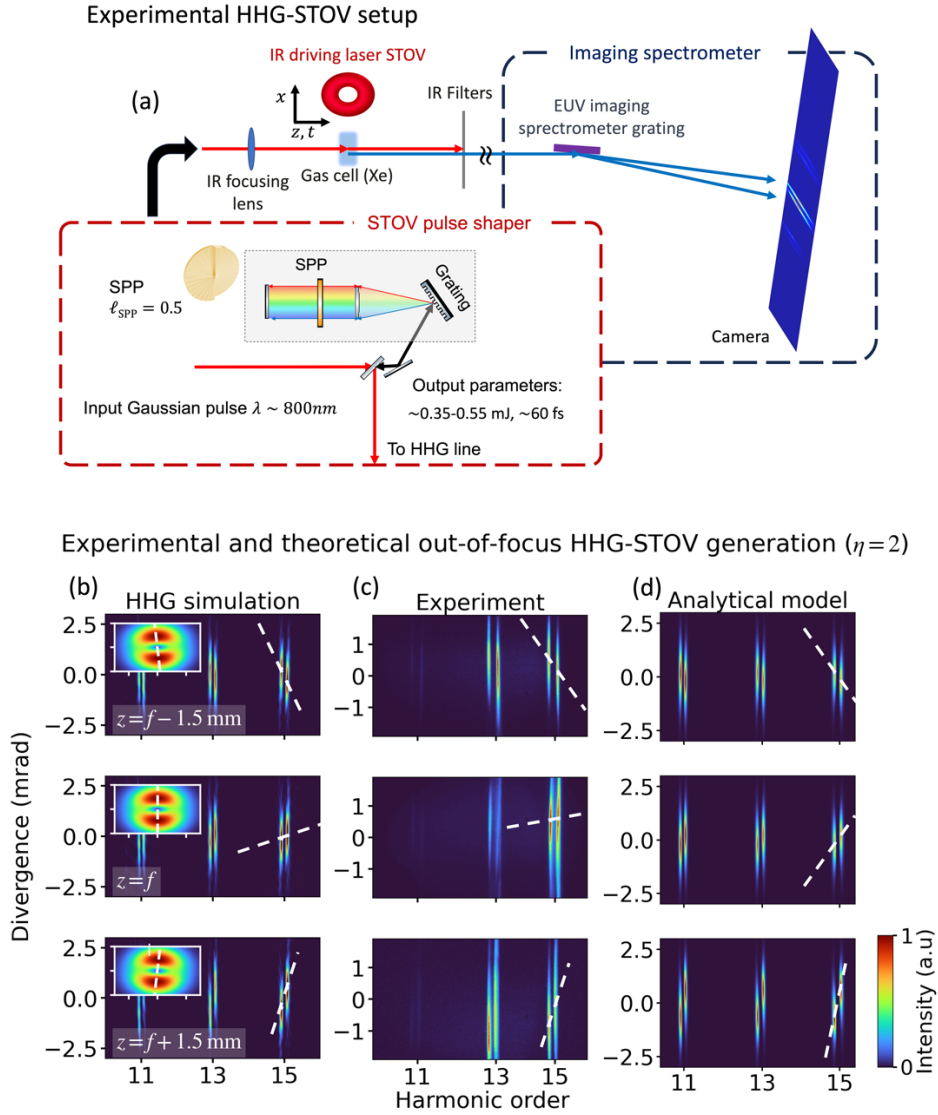

Figure 4: (a) Experimental setup for the generation of EUV STOVs via HHG. The driving Gaussian IR (800 nm) laser is sent to a STOV pulse shaper. In the pulse shaper, the Gaussian beam is spatially chirped using a reflective diffraction grating, collimated using a cylindrical lens and the spatiospectral phase is imprinted by a two-way round pass through the spiral phase plate (SPP), generating an IR SSOV. Finally, the IR SSOV reflects back in the diffraction grating, compensating the spatial chirp. The output from the pulse shaper is sent to the HHG line and focused into a STOV at the Xenon gas cell Finally, the EUV is directed to the measurement step, while the propagation of the remanent IR signal is blocked using a metallic filter. (b)-(d) Experimental and theoretical results of HHG driven by elliptical STOVs for different positions between the gas target and the focus. Top: before focus at $z = f-1.5mm$, middle: at focus $z = f$, and bottom: after focus $z = f+1.5mm$. HHG spectra peaks are labelled by white dots. (b) Full quantum HHG simulation spectra at different gas jet positions respect to the STOV focus. The driving STOV fields ($\eta = 2$) used are shown in the insets in each case with white dashed lines indicating intensity profile tilt. (c) Experimental harmonic STOV spectra results for an inhomogeneous driving STOV beam ($\eta = 1.8$) for the gas cell positions in panel (b). (d) Analytical model prediction of the strong field approximation for the far field harmonic spectra using the driving fields ($\eta = 2$) in panel (b) (see Methods).

It is interesting to note that, in the out-of-focus simulations the stripes edges for higher harmonic orders present an intensity warping at high divergences, as seen in Supplementary Information. The nature of this warping comes from the effect of the intrinsic dipole phase, which is more noticeable for the harmonics near the cut-off frequency.

Higher-order EUV STOVs and SSOVs can be obtained through HHG by using driving IR STOVs with higher topological charges. In the Supplemental Information we show the generation of highly charged EUV STOVs and SSOVs using drivers with $l_0 = 2, 4$. Remarkably, the bandwidth of the harmonics SSOVs is directly related to its topological charge, which allows us to confirm the generation of harmonic STOVs with $l$ up to 60 for the 15th harmonic.

# 3    Discussion and Conclusions

Generating STOVs presents significant challenges, as these complex beams are not natural eigenmodes of propagation. Up to now, the generation of STOVs was restricted to the IR/visible spectral regimes, and in most cases, it was limited to low topological charges. Recently, highly charged IR STOVs were observed using a spatial light modulator [14]. Our research, however, opens a pathway to generating high-frequency, short-wavelength, STOVs and SSOVs in a controlled manner. The exceptional coherence of HHG provides a unique capability to shape the spatiotemporal and spatiospectral properties of EUV light beams through an up-conversion process. We demonstrate, both theoretically and experimentally, that STOV up-conversion is feasible via HHG, despite the challenges posed by complex nonlinear dynamics and phase-matching requirements. This breakthrough enables the production of high-frequency light beams intrinsically coupling the smallest spatial and fastest temporal domains at which coherent light can be generated.

Specifically, we have shown that HHG can produce EUV light beams with intrinsic spatiotemporal coupling at nanometer and attosecond scales, characterized by a high transverse topological charge. We note that while the size of the focused EUV STOV beams is on the order of microns, techniques such as coherent diffraction imaging achieve spatial resolutions far below the beam size, thus potentially enabling nanometer-scale light matter interactions with structured light beams – for example, sub-wavelength (~10 nanometer resolution) EUV imaging has been achieved using micron-scale illumination beams [63]. Our findings, validated by careful comparison of analytical and numerical theories and experimental data, demonstrate the generation of both high topological charge EUV STOVs carrying transverse OAM and their conjugated counterparts, spatiospectral optical vortices (SSOVs), through HHG. We characterized the near- and far-field properties of these vortices, essential for understanding their propagation behaviour. Unlike HHG driven by longitudinal vortex beams, generating EUV STOVs is highly sensitive to several experimental conditions. As such, we show that the inhomogeneity parameter describing non-elliptical STOVs significantly influences the formation of spatiotemporal and spatiospectral EUV vortices.

The characterization of EUV STOVs and SSOVs, though being a remaining challenge, is addressed, by harnessing their intrinsic properties. Note that established EUV characterization techniques—such as attosecond beating by interference of two-photon transitions (RABBIT) or attosecond streaking—are inherently spatially integrated, making the complete spatiotemporal retrieval of EUV STOVs unfeasible. However, we demonstrate that the EUV STOV topological charge is related to its spectral bandwidth, therefore high topological charges can be characterized experimentally. This intrinsic property of STOVs circumvents current limitations in EUV spatiotemporal phase retrieval techniques, which are not yet developed for complete angular momentum characterization of high-order STOVs and SSOVs.

Our research paves the way for using STOVs and SSOVs at short wavelengths, extending to the EUV/soft x-ray spectral regimes, with controlled high topological charges. These beams are well-suited for probing ultrafast electronic dynamics—from femtoseconds to attoseconds—in systems with coupled spatial, temporal and spectral responses near absorption edges in the EUV/soft x-ray spectral regimes. Over the past decade, magnetic circular dichroism using high harmonic sources has provided unprecedented insights into ultrafast dynamics in magnetic materials [64,65]. Recent advances have shown that not only the spin, but the OAM of high-frequency light beams can be used to explore ultrafast dynamics in magnetic structures via the so-called magnetic helicoidal dichroism [66]. In this context, EUV/soft x-ray STOVs and SSOVs present a unique tool that, upon interaction with matter, can provide insights beyond those accessible with magnetic circular and helicoidal dichroism studies. In addition, the generation of structured EUV beams, with controlled OAM properties, is enabling unprecedented advances. For example, it has been proven that highly-divergent EUV beams with a vortex structure can achieve high-fidelity coherent imaging of highly periodic structures [46], which is not possible with standard Gaussian beams. These capabilities open a new dimension for investigating ultrafast light-matter interactions near the absorption edges of novel magnetic and quantum materials and for advancing the next generation of EUV coherent imaging techniques.

# Methods

## 4.1 Experimental setup

The generation of STOV driving beam to drive a nonlinear process shares the similar methodology as that in reference [27]. Nearly Fourier-transform limited (FTL, ~50 fs full-width half-maximum) optical pulses at a central wavelength of ~790nm were generated from a regenerative Ti:sapphire amplifier at 1-kHz repetition rate (KMLabs Wyvern HE). These FTL pulses were sent to a custom 4-f pulse shaper for STOV generation, which consists of a reflective grating (600 grooves/mm), a cylindrical lens (f=20 cm), a multi-faceted spiral phase plate (SPP, HoloOr, 16 steps per phase wrap) and a high-reflectance end mirror for retroreflection. To generate fundamental STOV pulses of spatiotemporal topological charge $l = 1$, a SPP with a designed spatial topological charge $l = 0.5$ at the design wavelength of 790 nm was used. Upon retroreflection from the end mirror, the pulse passed through the SPP twice and generated the desired spatial topological charge $l = 0.5 - (-0.5) = 1$. The pulse shaper serves as a 1-d Fourier transform in the $(\omega, t)$ domain, which converts the input FTL pulse to an SSOV pulse. Another spherical lens (f=25 cm) was used after the pulse shaper to focus the beam for HHG, which also performed as the Fourier transform in the spatial domain, converting SSOV pulses to STOV pulses at the focal plane.

For HHG, the driving pulse energy was set to 300-500 µJ impinging into a Xenon gas jet. The Xenon gas jet can be translated along the beam propagation direction to achieve HHG driven by the STOV beam at, before, or after the focal plane. The generated harmonics were transmitted through two 200-nm-thick aluminium filters (Lebow), which serves to block the residual driving beam. The characterization of HHG STOV beams shares a similar methodology as that in reference [20]. Instead of a direct spatiotemporal measurement, a spatiospectral measurement was performed to resolve the $(x, \omega)$ information of each harmonic orders. Although the spatiospectral phase wasn't collected for a comprehensive characterization of spatiotemporal light field, some major properties of STOV beam can still be extracted from the spatiospectral intensity [20], and such results can be compared and supported with our STOV HHG simulation. In our experiment, the spatiospectral data of harmonics were characterized by an imaging spectrometer (Hettrick), where the grating orders were aligned with the trivial dimension of the STOV beam. An EUV CCD camera (Andor Newton 940) was placed at the imaging plane of the spectrometer to record harmonics. The imaging spectrometer was placed at the far field after harmonic generation due to the technical difficulty to measure STOV beam at the gas jet. The separation between the gas jet and the spectrometer grating is sufficient (> 2m) such that (1) the approximation of far field propagation is valid, (2) the spatiospectral profile of STOV beam can be resolved by the camera.

## 4.2 Analytical model of HHG driven by STOVs.

The analytical model used at the beginning of Subsection 2.1 and in Figs. 3 and 4 is based on the strong field approximation [57]. If $E_0^{nf}$ is the driving near field, the $q$-the harmonic is described by $E_q^{nf} = |E_0^{nf}|^p e^{i(q\phi_0 + \phi_{int})}$, where $|E_0^{nf}|$ and $\phi_0$ are the amplitude and phase of the driving field. The $p$-th power corresponds to the HHG process non-perturbative intensity scaling, which is chosen to be $p \simeq 3$ (other close values give similar results), and $\phi_{int}$ is the intrinsic dipole phase [42,44]. The intrinsic phase has a posteriori been verified from the full simulations and experimental results to play a minor role with the driving STOVs, and therefore is neglected here. The harmonics $E_q$ are then propagated up to the far field using Fresnel-Fraunhofer diffraction integral:

$$E_{q,far}(\theta_x, \theta_y, t') = \frac{k_q}{2\pi i z} e^{\frac{ik_q(x^2+y^2)}{2z}} \int_{-\infty}^{\infty} E_q(x', y', t') e^{-ik_q(x'\theta_x + y'\theta_y)} dx' dy' e^{-i\omega_q t'}, \quad (5)$$

where $k_q = qk_0$ and $\omega_q = q\omega_0$, and then Fourier transformed in time to obtain the far field spatiospectra shown in Figs. 3 (c) and 4 (c).

### 4.3    Theoretical simulations of HHG driven by STOVs.

The HHG simulations are performed using a well-established framework, computing the full quantum single atom response and its macroscopic far-field propagation [62]. On one hand, the single atom response is obtained from the dipole acceleration calculated through the strong-field approximation, which presents a great accuracy vs computational time balance compared to the full Schrödinger equation numerical resolution. On the other hand, the gas target is discretized in individual radiators, and the single-atom response is calculated in each of them. The harmonic far-field emission is obtained as the collective propagation of the individual emissions using the electromagnetic field propagator [62], thus considering the high-order harmonics phase matching. This method has been successfully used to predict and understand several HHG experiments driven by structured light fields (see for example [39,40,44–46,61]). We have assumed an infinitesimally thin gas jet geometry, with a peak pressure of 5 Torr. In this bidimensional geometry only the transverse phase matching is considered, a valid approximation for low density gas jets as already reported in previous theoretical and experimental works [39,42,44].

The IR driving electric field used in the simulations is given by the spatiotemporal distribution in Eq. (2) describing the focusing dynamics of a spatiotemporal THL field of order $l$ in Eq. (1) at the lens of focal length $f$ = 25 cm, where $\sigma_x = \sigma_y = 1\ mm$ $1/e$ spatial widths in the $x$ and $y$ directions and $\sigma_t = 52$ fs temporal width. To avoid further numerical noise, the temporal envelope is modeled using a 64-cycle total-duration $\sin^2$ envelope, matching the 52 fs FWHM temporal width, instead the Gaussian temporal envelope. In all the simulations 800 nm driving wavelength, $1.6 \times 10^{14}$ W/cm$^2$ IR driver peak intensity and an atomic hydrogen gas target is assumed. Note that the HHG numerical simulations were performed in atomic hydrogen to reduce computational time. However, we do not expect any fundamental deviation in the results when compared to a xenon target (as in the experiment), which exhibits a similar ionization potential (13.6 eV in atomic hydrogen and 12.1 eV in xenon).

## Data Availability

Simulation and experimental data that support the findings of this study are available from the corresponding authors upon reasonable request.

## Code Availability

The simulation, experimental data and data visualization codes that support the plots and data analysis within this paper are available from the corresponding authors on reasonable request.

## Acknowledgment

This project has received funding from the European Research Council (ERC) under the European Union's Horizon 2020 research and innovation programme (grant agreement No 851201). C.H.-G. and L.P. acknowledge funding from Ministerio de Ciencia e Innovación (Grant PID2022-142340NB-I00). C.T.L. acknowledges support from AFOSR YIP (FA9550-23-1-0234) and partial support by DOE BER (DE-SC0023314). M.A.P. acknowledges funding from Ministerio de Ciencia e Innovación (Grant PID2021-122711NB-C21).

## Author contributions

C.H.-G., M.A.P., and C.-T.L. conceived the project. R. M.-H, M.A.P. and C.H.-G. performed the theoretical derivations and simulations and analyzed the resulting data. C.-T.L. and G.G. designed the experiment. G.G. conducted the experiment and analyzed the experimental data. C.H.-G., M.A.P. and L.P. supervised the theoretical simulations. M.M.M. and H.C.K. developed the facilities and measurement capabilities. C.-T.L. and M.M.M. supervised the experimental work. R.M.-H., G.G., L.P., C.-T.L., M.A.P. and C.H.-G. wrote and prepared the manuscript. All authors provided constructive improvements and feedback to this work. R.M.-H., L.P. and C.H.-G. Thankfully acknowledge RES resources provided by BSC in MareNostrum 4, and SCAYLE in Caléndula to RES-FI-2023-3-0045.

## Competing interests

H.C.K. has a co-affiliation as CTO of KMLabs Inc. M.M.M. and H.C.K. have a financial interest in KMLabs. The other authors declare no other competing interests.

## Corresponding Authors

Please address all correspondence to Chen-Ting Liao and Rodrigo Martín-Hernández

# Supplementary information

## S1 Higher topological charge STOV driving beams.

HHG allows for the generation of STOVs with extremely high topological charges, as indicated by the harmonic topological charge linear dependence with the driving STOV fundamental topological charge and the harmonic order. In the same way as in longitudinal vortices, where topological charges as high as 100 have been demonstrated [1] extremely high-topological charge EUV-STOVs can be obtained using multiply charged driving IR STOVs. Here, first we present HHG experimental results driven with higher topological charge STOVs with an inhomogeneous distribution and in the out-of-focus scenario.

In Fig. S1 we present experimental and numerical results, for HHG driven by IR STOVs with topological charges of $l_0 = 1$ (first row), $l_0 = 2$ (second row) and $l_0 = 4$ (third row). In this scenario, for all the cases an inhomogeneity parameter $\eta \sim 2$ is assumed, and the gas jet is located after the focus. As we are dealing with an out-of focus and inhomogeneous scenario, the positively sloped spectral stripes, discussed in the Section 2.3, for each harmonic order are observed. Looking at the experimental spectra [Fig. S1(c)], the increase of the driving topological charge enlarges the spectral distance between the two stripes for each harmonic order, i.e., their harmonic bandwidth is increased. As there is a direct relation between the topological charge content and the spectral bandwidth in STOVs, for a given harmonic order, a larger bandwidth means a higher topological charge, which linearly scales with the harmonic order and the fundamental topological charge.

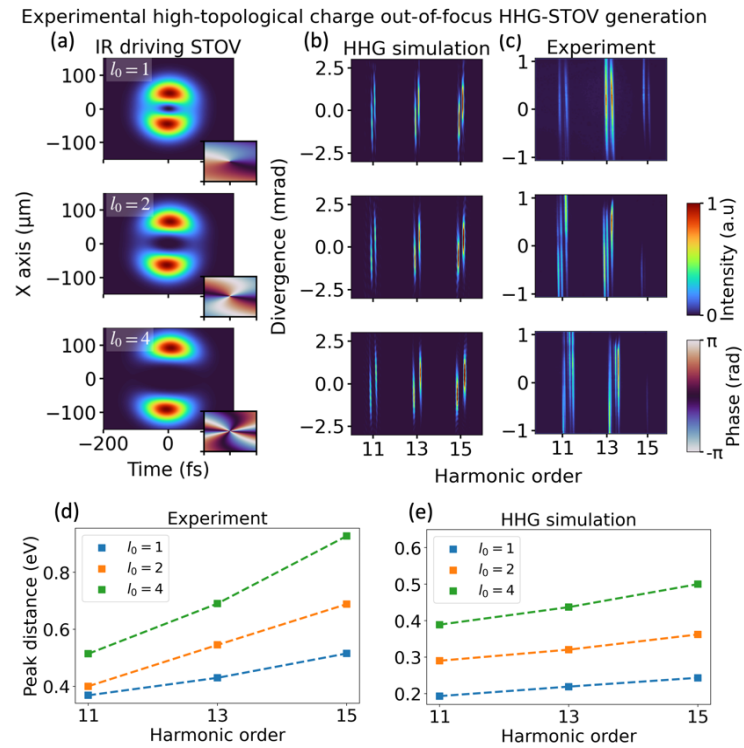

Figure S1: Experimental and theoretical results of HHG driven by elliptical STOVs for different topological charges after-the-focus position. Top: charge $l0 = 1$, middle: $l0 = 2$, and bottom: $l0 = 4$. (a) The driving STOV fields ($\eta = 2$) with the different topological charges used. (b) Simulated HHG-STOV spectra results for the driving STOV fields shown in panel (a). (c) Experimental harmonic STOV spectra driven by an inhomogeneous STOV beam ($\eta \sim 1.8$) for the different topological charges. Note the positively slope striped harmonic spectra resulting from the inhomogeneous driving STOV in the after-focus-position. The spectral peak distances of the harmonic stripes are shown in panels (d) and (e) for the experimental and simulated results presented in panels (b) and (c). The predicted harmonic bandwidth increasing with the driving beam topological charge is experimentally confirmed.

The theoretical simulations, shown in Fig. S1(b), are performed feeding the driving spatiotemporal intensities shown in Fig. S1(a) and assuming a gas jet position $z = f + 1.5$ mm to mimic the experimental parameters. The simulation spectra show a very similar trend, predicting the harmonic bandwidth widens as the driving topological charge is also increased. The peak distances between the harmonic stripes in Figs. S1 (b) and S1(c) are shown in Figs. S1(d) and S1(e) as function of the driving STOV topological charge, respectively. It is evident the direct relationship between the driving STOV topological and the increasing harmonic STOV bandwidth. Although a complete spatiotemporal/spatiospectral phase retrieval would be required to determine the exact harmonic STOV/SSOV topological charge—which to the best of our knowledge does not exist in the EUV regime—, this is clear signature of the higher harmonic topological charges. The scale difference between experimental and the simulated results comes from the different driving pulse durations assumed in the numerical simulations due to computational limitations.

## S2    Intrinsic dipole phase effect in the high-order harmonic STOV

The intrinsic dipole phase is the accumulated electronic phase during its continuum excursion which is translated into the HHG emission in the recombination step. It is known to be critical on several applications of the HHG emission, as the harmonic beam divergence optimization, the generation of attosecond pulses or the intra-atomic temporal optimization [2–4]. We have observed that the perfect agreement between the HHG numerical simulations and the analytical model (which does not include the dipole phase) shown in the main text hold when the harmonics are well in the plateau. However, we do observe dipole phase effects for the harmonics closer to cut-off energy.

To analyze the effect of the dipole phase, in Fig. S2, SSOV higher harmonic orders are shown (from the 21st to the 25th), closer to the cut-off ($q_{cutoff} = 27$). The spectrum is obtained assuming that the gas-jet is placed in the focus position and using a driving STOV inhomogeneity parameter $\eta = 2$. The results shown are obtained using (a) the full numerical simulations, and (b, c) the analytical model shown in Eq. (5) neglecting/including the intrinsic dipole phase $\phi_{int}$. The intrinsic dipole phase is calculated using the gamma model [2]. In contrast with the straight harmonic stripes obtained in Fig. 3, bottom row, the higher harmonic orders shown in Fig. S2 show their edges warped.  The nature of the harmonic warping obtained in the full HHG numerical simulations can only be described when the intrinsic dipole phase term is included in the analytical model.

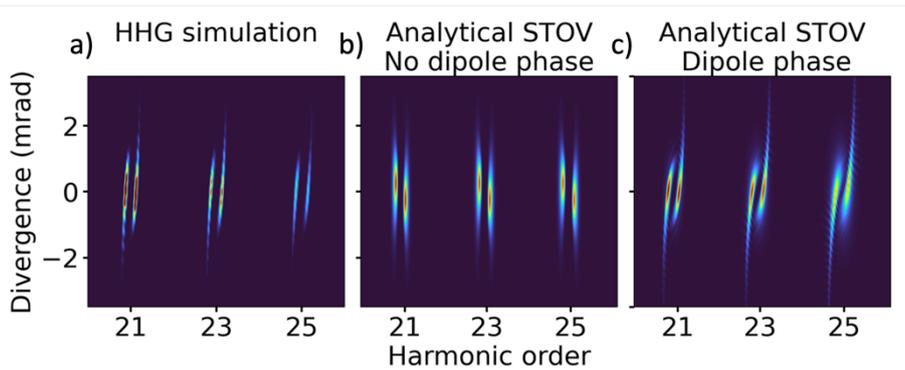

Figure S2: Intrinsic dipole phase effect on the inhomogeneous harmonic SSOV spectrum. (a) High-order harmonics (near to cutoff frequency) driven by an inhomogeneous STOV ($\eta = 2$). The harmonic stripes are bent respect to the lower harmonic orders (see Fig. 3 bottom row in main text). The analytical model harmonic spectrum (see methods) without the intrinsic dipole phase (b) and with the intrinsic dipole phase (c). Only when the dipole phase term is included, the harmonic SSOV bending can be described

## S3 Elliptical STOV beam generation from the pulse shaper

The ellipticity of STOV beam is defined by the inhomogeneity parameter $\eta$, as described in eq. (3). The $\eta = 1$ case represents the canonical STOV beam, where a perfect donut-shaped STOV beam is formed. In many cases, the inhomogeneity parameter can be deviated from unity due to various experimental conditions. In our experiment, such inhomogeneity parameter $\eta$ is defined by the beam spatial ellipticity at the Fourier plane of the 4-f pulse shaper. The inhomogeneity parameter $\eta = 1.8$ can be calculated by using the following experimental conditions, i.e., beam diameter (10 mm), laser spectral bandwidth (45 nm), grating groove density (600 grooves/mm), cylindrical lens focal length (20cm).

Figure S3 (a) shows the simulation of an elliptical Gaussian beam at the Fourier plane of the pulse shaper. The STOV beam at the gas jet can be estimated with a 2-d Fourier transform, as shown in Figure S3 (b), where spatiotemporal profile of the STOV beam is very similar to our simulation in the paper with inhomogeneity parameter $\eta = 2$.

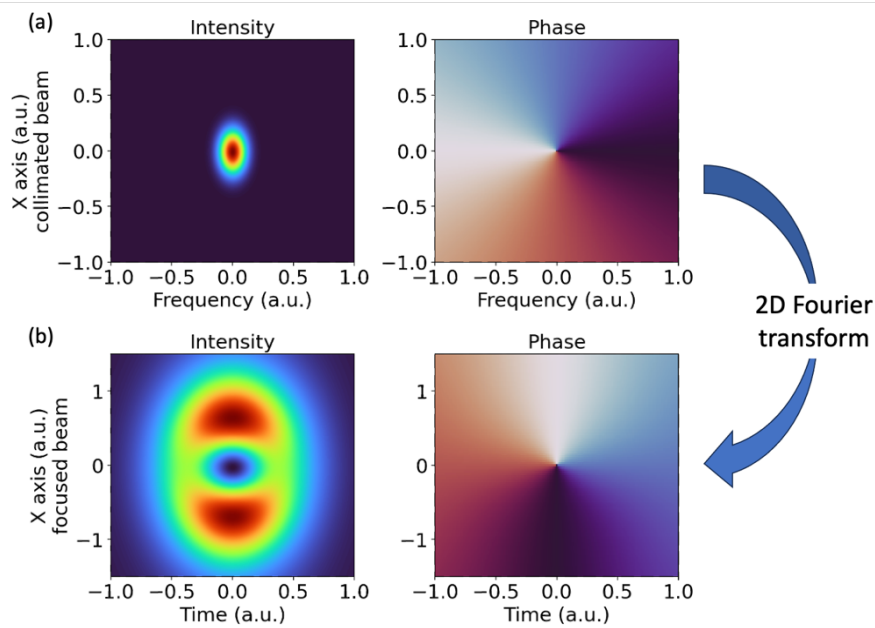

Figure S3: Simulation of STOV beam generation from pulse shaper with inhomogeneity parameter $\eta = 1.8$. (a) An elliptical Gaussian beam is formed at the Fourier plane of the pulse shaper. (b) The Elliptical STOV beam at the gas jet generated from the (a).